# Who Benefits? Employer Subsidization of Reproductive Healthcare and Implications for Reproductive Justice

*Annie McGrew, University of Massachusetts Amherst*

*Yana van der Meulen Rodgers, Rutgers University*

**Abstract:** With the reversal of *Roe v. Wade* in 2022, many U.S. employers announced they would reimburse employees for abortion-related travel expenses. This action complements increasingly common employer policies subsidizing employee access to assisted reproductive technologies such as in-vitro fertilization and egg freezing. This article reflects on why employers offer these benefits and whether they enhance or undermine reproductive justice. From the employer's perspective, abortion and assisted reproductive technologies help women to plan childbearing around the demands of their jobs. Both are associated with delayed childbirth and reduced fertility, which lower the costs of motherhood to employers. However, firm subsidization of these services does not further reproductive justice because it reifies structures which incentivize women to delay childbirth and reduce fertility, and it reinforces economic and reproductive inequalities. We conclude by questioning whether reproductive justice is possible without transforming the economy so that it prioritizes care over profits.

**Acknowledgments:** The authors thank Nancy Folbre, Kelly Jones, Lenore Palladino, Rishita Nandagiri, Leigh Senderowicz, Wendy Sigle, Angelina Chapin, Katherine Moos, Randy Albelda, Mieke Meurs, two anonymous reviewers, and participants at the 2022 IAFFE, 2023 ICAPE, 2023 EEA, 2023 AHE, and 2024 ASSA conferences for their helpful comments.

## 1. Introduction

The U.S. Supreme Court's overturning of *Roe v. Wade* in June 2022 left abortion legality up to the states, resulting in complete bans or very restrictive abortion laws in about half the states. This major legal challenge to women's reproductive rights prompted numerous corporations to announce that they would cover travel costs for employees needing to go out of state to obtain an abortion. Subsidization of abortion-related travel costs is the latest addition to a series of reproductive healthcare benefits that employers have started to offer, especially employer-provided health insurance covering assisted reproductive technologies such as in-vitro fertilization (IVF), egg freezing, and surrogacy.

Employer support for assisted reproductive technologies has garnered much attention in media and academic discourse, with some controversy. Some argue that these benefits empower women by giving them control over their fertility, freeing them from the biological constraints associated with age-related fertility decline and allowing them to pursue career advancement and financial stability before having children (Varlas et al. 2021; Mattson 2017). However, others argue that society should make it easier for women to have children earlier in their careers so they don't have to resort to technological fixes that are economically, emotionally, and physically costly. Egg freezing in particular is seen as a temporary, technological band-aid for a structural problem because it falls short of solving a deeper problem, namely that workplace norms are molded around the life cycle of men (Waldby 2014, Geisser 2018, Harwood 2023, O'Rourke et al. 2023). Our paper contributes to this debate, which has largely been conducted in the context of egg freezing, by broadening the range of reproductive health benefits offered by employers and also asking why these policies might benefit employers. In addition, we integrate a reproductive justice lens and argue that these policies fail to address the core problem: the contradiction between production and reproduction which makes childrearing costly. Moreover, they exacerbate economic and reproductive inequalities, thereby reinforcing labor markets as gendered institutions.

Following in the tradition of reproductive justice, we do not assume that abortion and assisted reproductive technologies are inherently empowering technologies. Rather, we evaluate the social and economic systems in which reproductive health decisions are made and how these systems interact with employer subsidization of reproductive healthcare in a way that furthers or detracts from reproductive justice. Our analysis places women within an economic context which privileges production over reproduction, putting them in positions where they must forsake one for the other. Because women bear the primary responsibility of reproduction (biologically and socially), and because labor markets are "bearers of gender," women bear most of the economic costs of motherhood (Elson 1999).

Employers must also grapple with the contradiction between production and reproduction as motherhood is costly for them. When women have children, many take leave, work fewer hours, switch companies, or leave the labor force altogether, all of which entail turnover and productivity costs for employers. These costs reflect the loss of control that employers exercise over women workers after childbirth and the resulting difficulty that employers have extracting surplus labor from their workers. Because abortion and assisted reproductive technologies are associated with delayed childbirth and reduced fertility, employer subsidization of these technologies reduces the (present) costs of motherhood to employers. Employers also may benefit from providing these benefits because they increase the potential to hire and retain women workers and they strengthen signaling values to stakeholders. Thus, employers may view these policies as good for business.

These reproductive healthcare benefits do not necessarily further reproductive justice because they do little to ease the contradiction between production and reproduction. Women are still in the position of having to sacrifice economic goals for their reproductive goals, and vice versa. Although women may "choose" to manage the contradiction between production and reproduction by delaying childbirth, the high economic cost of childrearing, especially early in their careers, raises questions about how much agency women have in their reproductive choices. Without easing the economic strain on women of childrearing, employers give women the illusion of choice by subsidizing fertility-regulating reproductive services.

Employer subsidization of reproductive healthcare exemplifies the way in which reproduction is stratified along social hierarchies, as high-income workers are most likely to have access to these benefits. For high-income women, provision of these benefits may increase the pressure on women to delay childbirth to avoid being put on the "mommy track."[1] For low-income women who have fewer resources to support their reproduction and who receive little support from the state, working at a company offering support for assisted reproductive technologies may be the only way to get access.

The remainder of this paper is structured as follows. Section 2 sets the context with a brief background on employer support for abortion-related travel costs and assisted reproductive technologies, as well as an overview of the reproductive justice framework. The subsequent three sections present our analysis of why employers benefit from providing these reproductive health benefits: section 3 posits that motherhood is costly for employers due in large part to replacement costs associated with a reduction in women's labor supply; section 4 shows why employer support for reproductive health services can reduce these costs of motherhood; and section 5 presents other mechanisms through which employers may benefit from subsidizing abortion-related travel costs and assisted reproductive technologies. Thereafter we examine the costs of motherhood to women as well as the social and economic context in which women are situated. In particular, section 6 discusses how women face a double bind in which they must choose between two equally dissatisfactory outcomes; and section 7 describes the stratified ability of people with different social identities to control their reproduction, and how corporate support for assisted reproductive technologies exacerbates these inequalities. The final section considers how policies situated within a reproductive justice framework can help to alleviate the double bind.

---

[1] We use the labels high-income and low-income to distinguish between women with high educational attachment who work in high-wage jobs which offer a range of benefits, and women with lower educational attainment who work in lower-wage jobs with fewer benefits. High-income women are presumed to occupy a higher status position in society, allowing them greater access to resources to finance their reproductive healthcare needs. Low-income women are presumed to be working class, less privileged, and more constrained in their options for financing their healthcare needs.

## 2. Background

### a. Employer Support for Abortion and Assisted Reproductive Technologies

Since abortion first became legal in the U.S. in the 1973 landmark Supreme Court decision, *Roe v. Wade*, conservative groups and politicians have attacked women's access to abortion, resulting in a series of state-level regulations affecting both abortion seekers and providers. These restrictions have resulted in clinic closures, fewer available appointments, and longer travel times and distances to obtain an abortion. The restrictions have also increased the monetary costs associated with abortion, which is already an expensive procedure and relatively difficult to finance for low-income women (Fuentes et al. 2016, Lindo and Pineda-Torres 2021).

In June 2022, the U.S. Supreme Court overturned *Roe v. Wade* in the case *Dobbs v. Jackson Women's Health Organization*. This ruling left the legality of abortion up to the states, with abortion being completely banned in 14 states and banned at an early gestational age in another 2 states as of February 2024.[2] With the banned states clustered in the South and Midwest, many women have needed to travel long distances to access abortion services (Rader et al. 2022). This increase in travel distances prompted numerous companies to announce they would assist their employees in paying for some or all of their employees' abortion-related travel expenses.[3] Some of the first companies to announce this policy included well-known corporations such as Starbucks, Tesla, Yelp, Airbnb, Microsoft, Netflix, Patagonia, JPMorgan Chase, Levi Strauss, and PayPal. Figure 1 presents a set of word cloud diagrams depicting the names of companies with at least 500 employees that announced some kind of support for abortion access around the time of the Dobbs decision. The company names are arranged by sector and are depicted according to company size, as measured by number of employees. A complete list of the benefits is found in the Online Appendix.

Many of the same employers offering reimbursement for abortion-related travel also provide coverage for assisted reproductive technologies in their employer-provided health insurance plans. These benefits gained traction in the U.S. when Apple and Facebook announced in 2014 that they would cover egg freezing for their employees (Tran 2014). In the past few years, financial support for assisted reproductive technologies for employees has become increasingly common, especially at large companies. In the U.S., almost 30% of employers with 500 or more employees provided IVF benefits for their employees and 11% covered egg freezing in 2020. Just five years earlier, only 23% of employers covered IVF and 5% covered egg freezing (Dowling 2021).

---

[2] Several organizations have abortion law trackers providing this information, including the *New York Times*, Guttmacher Institute, and Kaiser Family Foundation.

[3] Other support for abortion includes communication affirming abortion access and paid time off for abortion travel and recovery. Unfortunately, we do not know the uptake of these benefits, nor do we have information about the confidentiality around using these benefits. Filing for abortion-related travel costs could potentially involve additional filings with insurance and HR systems, so employees might be reluctant to take advantage of the benefits.

**b . Reproductive Justice**

Reproductive justice as a concept was founded by a group of Black women in the 1990s in response to the prevailing paradigm which focused on reproductive health technologies enabling women to avoid childbirth without addressing the systemic barriers that prevented women from being able to have and raise children in safe conditions. Reproductive justice advocates argued that an individual's ability to have a child under conditions of one's own choosing and to be able to parent that child in healthy environments was crucial for women's dignity and reproductive autonomy (Ross and Solinger 2017). Advocates thus questioned the extent to which contraception and abortion benefited women if economic or social constraints prevented them from being able to have children and parent those children in safe and healthy environments. If the environment is left wanting, women lack meaningful choice in whether and under what conditions to have children. The reproductive justice framework thus focuses on the ways in which economic and social systems constrain the options of individuals.

The questioning of choice shifts the focus of analysis away from individuals towards system and institutions. Thus, an analysis rooted in reproductive justice must begin with an understanding that choice and context are inextricably linked – that the context in which women make reproductive decisions is integral to understanding how much agency they really have in their choices. For example, without the resources to be able to have children, the decision not to have a child may be more about survival than choice. For some women, having a child can induce poverty (Foster et al. 2022). Hence socioeconomic concerns are an important motivation for women to obtain abortions across many countries (Chae et al. 2017, Finer et al. 2005).

An analysis grounded in reproductive justice aims not to shift individual behaviors to adapt to unjust conditions, but rather to identify, challenge, and transform the unjust conditions in the first place. Rather than argue for access to abortion as a means to alleviate poverty for women, reproductive justice invites us to imagine and work towards a system in which an unintended pregnancy does not lead to economic ruin. In this vein, Ross and Solinger (2017:161) propose a well-grounded alternative:

> "If…the person who became pregnant unintentionally had access to comprehensive reproductive health services and time off to care for herself and her family…she might be able to manage the pregnancy. If the pregnant person had a stable work-shift schedule, denied most often to the lowest-paid workers, and if she had access to child care, a pregnancy might not spell economic devastation."

Our analysis is written in this spirit. Rather than evaluate how employer support for abortion travel and assisted reproductive technologies affect individual women, we focus on whether these policies challenge, uphold, or exacerbate the systemic barriers which prevent women from achieving reproductive justice. Moreover, we explore the ways in which the capitalist system, by prioritizing production, penalizes reproductive labor and in doing so, creates incentives for employers to offer benefits that help them to reduce the costs of reproduction.

## 3. Costs to Employers of Motherhood

The cost of motherhood to employers (and to women) is one manifestation of the contradiction between production and reproduction in today's economic system. In this section, we provide evidence indicating that the costs of motherhood (one measure of the cost of reproduction) are high for employers and, in the next sections, that fertility-regulating technologies may help employers reduce those costs in a way that doesn't address their source. Thus, these policies help employers manage the contradiction between production and reproduction without easing the contradiction.

There is ample evidence that motherhood constitutes a shock to women's labor market outcomes in ways that are costly to employers. When women become mothers, they often become less attached to the labor force and their employer. Most women in the labor force take time off for childbirth, usually with some type of paid or unpaid leave. In the U.S., 42% of women take a paid leave after the birth of a first child, 31% take unpaid leave, 19% exit the labor force, and the remainder have some other arrangement (Goldin and Mitchell 2017). Over time, across different cohorts of U.S. women starting with those born in the late 1930s, labor force participation rates have dropped markedly after the birth of a first child (Goldin and Mitchell 2017). Moreover, women's attachment to the labor force is consistently lower for women with children under the age of 3, and for women with children under the age of 6, compared to women with school-aged children (Figure 1). These patterns highlight the ways in which increased care burdens disrupt women's attachment to the labor market.

Even when women decide to continue working after having children, they may move to a different company that better accommodates childrearing – a move that would entail a loss for the original company. For example, in Norway, mothers are more likely to change occupations and employers compared to women without children, and women who make such changes as a result of having a child are likely to switch into lower-paying jobs (Lundborg et al. 2017). This indicates that women may switch into jobs where they sacrifice higher wages for child-friendly benefits such as flexibility or shorter commute (Mas and Pallais 2017, Bertrand 2020).

Motherhood is clearly associated with worker turnover, which is costly to employers in terms of search costs, disruptions, and lost firm-specific human capital. Searching for, hiring, and training new workers costs both time and money, while increasing the hours worked of existing employees to substitute for new mothers who have quit or reduced their hours may involve paying overtime. Estimating these costs is challenging and depends on a variety of factors that include the industry, availability of substitutes, specificity of the human capital, and external labor market conditions. Focusing just on the cost of searching for and training a new worker may miss important features of turnover costs, including the increased salary costs of keeping incumbent workers whose value increases following the departure of a coworker. Turnover costs also include the additional kinds of firm-specific human capital that are not directly gained from training programs, as well as the cost of finding a replacement worker who is a strong match, which can be more difficult in thin labor markets.

In Germany, employers incurred large replacement costs of approximately two times the annual salary of a departed worker (Jäger and Heining 2022), and in Sweden, following the implementation of an extended leave policy, employers encountered an increase in their total wage bill due to worker turnover, with an average cost of 60% of the salary of a full-time worker going on extended leave (Ginja et al. 2023). These increases in the total wage bill in Germany and Sweden are large, and they fit within the range found in an earlier systematic review of the cost of employee turnover in the U.S. In particular, Boushey and Glynn (2012) found that the cost of replacing employees can amount to about 16% to 20% of the annual salary of a mid-level employee while replacing an executive-level employee can cost more than double her salary. Employers are likely to assume these costs when their employees have children. Employers may even decrease promotion, hiring, and wages in anticipation of these costs, an assertion supported with evidence in Ginja et al. (2023) showing that industries with higher exposure to the parental leave law had lower promotion rates, hiring rates, and starting wages for women of childbearing ages.[4] Employers thus appear to understand the costs of motherhood and adjust their behavior accordingly.

Motherhood can also be costly to employers in ways that are more difficult to measure. For example, motherhood reduces an employer's control over their employee's productivity, which makes mothers less than ideal workers. Because of social norms which ascribe to women the primary responsibility to care for their children, mothers may not be able to work long or odd hours or travel for work. Furthermore, motherhood and pregnancy make demands on the body such as lack of sleep that may reduce productivity in the workplace. As a result, employers exercise less control over mothers than workers who are less burdened by care responsibilities, making it more difficult for employers to extract surplus value from these workers.

Studies on the effect of having children on worker productivity are sparse. One exception is Gallen (2023), who finds that mothers are substantially less productive than non-mothers and men with and without children.[5] In addition, some evidence suggests that a productivity gap emerged between men and women academics during the pandemic because of the uneven distribution of care work, especially among assistant professors who are younger and more likely to have young children at home (Squazzoni et al. 2021). Evidence for other occupations such as financial analysts similarly suggests that working from home resulted in a decline in productivity for women with young children (Du 2020; Barber et al. 2021). Hence the increased care burden for women brought on by the pandemic (and the existing gendered distribution of care work) appear to have entailed productivity losses in the workplace.

---

[4] This point about discrimination against pregnant women and mothers is echoed in numerous studies, including Gallen (2019), Jessen et al. (2019), and Huebener et al. (2022).

[5] A possible explanation for this productivity gap is a flexibility penalty in which job structures that allow greater flexibility in hours worked are less productive, perhaps because flexible jobs tend to be jobs where measuring productivity is more difficult. When women work in jobs with flexible hours, they are about 40% less productive than men (Gallen 2023).

## 4. Reducing the Costs of Motherhood for Employers

We posit that employers benefit from offering their workers support for abortion-related travel and assisted reproductive technologies because these benefits help employers manage the contradiction between production and reproduction by reducing the costs of motherhood to the employer. Access to abortion and assisted reproductive technologies facilitate delayed childbirth and reduced fertility, both of which are cost-reducing for the employer. In the case of reduced fertility, employers avoid costs simply because fewer of their employees are having children. Additionally, delayed childbirth is associated with decreased fertility because the probability of a successful pregnancy decreases with age and because delays reduce the timespan for having more children (Balasch and Gratacós 2011; Tavares et al. 2016; Beaujouan et al. 2023; Beaujouan 2023). Delayed childbirth may also result in involuntary childlessness (Tavares et al. 2016), and it allows for downward revisions of childrearing intentions as other competing life objectives emerge (Tavares et al. 2016, Beaujouan 2023). The absence of supportive family-friendly policies has also been linked to delayed childbirth and lower fertility (Bratti 2023).

Even if delayed childbirth does not reduce fertility, employers may view delayed childbirth among their workers as cost reducing if they discount future returns heavily and instead focus on efforts to reduce their present costs. Delayed childbirth, then, reduces the present costs associated with worker turnover as it pushes the costs of motherhood into the future. Recent evidence on firms trading in U.S. public markets suggests that firms indeed have become more short-term oriented as investors are increasingly discounting the expected future cash flows of the companies in which they invest in favor of short-term returns (Sampson and Shi 2023). Delayed childbirth among workers also benefits employers because it maximizes the time women spend in the labor force over their whole life cycle (Goldin and Mitchell 2017), and is associated with higher labor market attachment (Bratti 2023). In addition, women who have children later in life are more likely to return to the same employer post-childbirth (as opposed to switching employers, becoming unemployed, or leaving the labor force), which increases their firm-specific human capital and limits turnover costs for employers (Sandler and Szembrot 2019).

If abortion and assisted reproductive technologies *increase the likelihood* that women delay childbirth or have lower fertility, then these policies can benefit employers. Indeed, increased access to abortion and assisted reproductive technologies reshapes women's life timing, leading to delayed age at first birth, delayed marriage, increased educational attainment, and more favorable labor market outcomes (Abramowiz 2014, Kroeger and La Mattina 2017, Ohinata 2011, Gershoni and Low 2021a,b, Myers 2017; Rodgers et al. 2020). Increased access to abortion and assisted reproduction technologies can increase the likelihood of delayed childbirth or reduced fertility in several ways. Assisted reproductive technologies, for example, affect women's life choices early on as they anticipate the availability of the technology before the realization of their family goals later in life. Thus, access expands women's expectations of their fertility horizons. Because of the high economic costs of childrearing, access to abortion and assisted reproductive technologies may encourage young women to delay childbirth (Gershoni and Low 2021a). On top of that, evidence suggests that women may be making fertility decisions with imperfect information about the links between delayed childbirth and infertility, increasing the likelihood of involuntary childlessness (Tough et al. 2007, Daniluk and Koert 2012).

Thus far we have argued that employers benefit from access to fertility-regulating reproductive healthcare because it reduces the costs associated with motherhood by delaying childbirth and reducing fertility. Employers should gain from women's access to contraception in a similar way as they do from women's access to abortion and assisted reproductive technologies. Access to contraception also reshapes women's life plans by delaying age at first birth and increasing lifetime labor force participation (Bailey 2006). Employer subsidization of contraception decreases costs to the employer by avoiding pregnancy-related expenses such as absenteeism, decreased productivity, employee loss, and paid leave (Canestaro et al. 2017). However, we do not include contraception in our analysis because the 2010 Affordable Care Act (ACA) mandated employer-sponsored insurance plans to cover contraception.[6] Thus, employers do not have the choice to offer it voluntarily anymore. The fact that employers subsidize abortion and assisted reproductive technologies voluntarily is suggestive that they benefit from these policies or, at least, that these policies do not hurt profits.

## 5. Other Benefits to Employers

Media articles covering employer subsidization of abortion and assisted reproductive technologies have framed employers' motivation in terms of attracting and keeping talent and increasing workforce diversity (Tran 2014; Goldberg 2022; Sorkin et al. 2022; Evers-Hillstrom 2022; Dowling 2021). Some sources have connected the increase in voluntary benefits offered by employers to an attempt to become more competitive amidst the tight post-pandemic labor market of the "Great Resignation" (Mayer 2021). This is supported by evidence from Adrjan et al. 2023 that employers announcing reimbursement for abortion-related travel expenses received more clicks on their online job postings after the announcement.

---

[6] Before the ACA, many states had their own mandates requiring employers to cover contraceptives. However, in states without a mandate, around half (47-61%) of employers still covered contraception (Sonfield et al. 2004).

Viewed within an imperfectly competitive labor market, however, amenities can generate labor market power. A growing body of research suggests that most labor markets are not perfectly competitive, but instead that search frictions and heterogeneity in worker preferences lead to monopsony power even when there are many employers in the labor market (Manning 2021). Thus, offering an amenity that is highly valued by women, could increase employer labor market power and their ability to set wages. A large literature in health and labor economics suggests that employer provided health insurance creates job lock, which describes the inability of a worker to leave a job for fear of losing employee benefits. Job lock is related to monopsony power because both derive from a worker's inability to move freely between jobs (Edwards 2022, Wang 2021). Employers may benefit from reduced turnover associated with job lock as well as increased wage-setting power associated with increased labor market power under imperfect competition.

Providing financial support for reproductive healthcare may also strengthen workers' attachment to the employer through loyalty: employees are thankful to the employer for providing this benefit when other employers do not, and then feel loyal or indebted to the employer (Roehling et al. 2001). Indeed, surveys conducted by Maven Clinic and Fertility IQ find that employees who received coverage for assisted reproductive technologies from their employer felt more loyal and committed to that employer, and that a majority of respondents would switch jobs if such benefits were offered (Maven Clinic 2023; FertilityIQ 2021). The Fertility IQ survey also found that women who had IVF treatment fully covered by their employer were more likely to return to that employer after maternity leave compared to those without support for assisted reproductive technologies.

Employers may also benefit by subsidizing abortion-related travel and assisted reproductive technologies through signaling their principles to consumers and other stakeholders. The literature on corporate social responsibility and activism suggests that companies which address social challenges can achieve equal or better financial performance because they are rewarded by their employees, customers, and other stakeholders. Although employers may gain financially from taking political stances that align with their stakeholders, they also may incur financial costs when choosing sides on divisive issues that do not align with stakeholders (Durney et al. 2020, Mkrtchyan et al. 2022, and Hou and Poliquin 2022, Bhagwat et al. 2020) Thus, taking a stance on political issues can be a "double-edged" sword: it can build loyalty among stakeholders, or it can alienate segments of the population (Chatterji and Toffel 2018, Larcker et al. 2018).

In the case of abortion, a highly divisive issue, companies may be reluctant to take a stand when they expect a value misalignment with customers. Survey data examined in Chatterji and Toffel (2018) and Larcker et al. (2018) indicate that while the majority of respondents view corporate activism on environmental and social issues in a positive light, less than 40% of respondents support corporate activism on abortion, and taking a political stance on abortion was ranked as less favorable than stances on racial issues and LGBTQ rights. Employers may face a tradeoff between appeasing current employees and attracting new ones when taking a political stand on abortion. In particular, Adrjan et al. (2023) find that employers announcing reimbursement for abortion-related travel expenses received more clicks on their online job postings after the announcement, but at the cost of lower job satisfaction ratings from their current employees. Hence taking a political stance on abortion may alienate more stakeholders than it attracts. Employers might also be influenced by employee groups who push their employers to offer these benefits. For example, after *Roe v. Wade* was overturned, workers at Google and Amazon shared a signed petition with their employer demanding that they support abortion rights with actions such as expanding travel benefits for abortion to all workers, denouncing the Supreme Court decision, and ending their donations to anti-abortion politicians (Lima 2022).

At a minimum, these policies allow employers to manage the contradictions between production and reproduction without addressing its root causes – profit as the lifeforce of the corporation. Delayed childbirth and reduced fertility are likely to close gender gaps in economic outcomes, allowing employers to demonstrate progress on gender inequality, but they do so by shifting the costs of reproduction onto women rather than solving the ultimate problem that motherhood is viewed as a cost rather than a benefit. Capitalism, thus, addresses the care crisis through care fixes which "resolve nothing definitely but merely displace the crisis, thereby perpetuating the structural reflex of capitalist economies to offload the cost of care to unpaid sectors of society" (Dowling 2021, pg. 15). In a similar way, delayed childbirth and reduced fertility displace the contradiction between production and reproduction onto women bodies and in doing so, avoids dealing with the structural conditions that led to the problem, which would be much more costly.

Finally, these policies allow for fixes that do not challenge, but rather reinforce, the gendered nature of labor markets. Feminist economists have long argued that labor markets are gendered institutions. Elson (1999), for example, argues that labor market institutions are "bearers of gender", meaning that gendered stereotypes are inscribed in social institutions. A case in point is front loading, a gendered workplace structure that excessively burdens employees to work very hard and succeed quickly in the first few years of their career. Historically, workplaces were built on gendered models that assumed the worker has a wife who does not work and takes care of the workers' needs outside of work. However, today with so many women working double shifts, both in the office and at home, front-loading creates a problem because it overlaps with women's relatively short window for childbearing. Employer subsidization of abortion and assisted reproductive technologies, thus, reinforces the gendered workplace structure by incentivizing women to change their biology rather than adjusting the workplace to support reproduction at any age (McGinley 2016).[7]

## 6. The Double Bind

Motherhood is not only costly for employers, but also for working women who, on average, experience a decline in earnings following the birth of the first child. This motherhood penalty constitutes a salient feature of labor markets around the globe. For example, women's average annual earnings losses within five to ten years after first birth range from 21% in Denmark to 61% in Germany, with mothers in the U.S. experiencing a 31% earnings loss (Kleven et al. 2019). These earnings losses – which are large, immediate, and persistent – are driven by some combination of a drop in employment, fewer hours worked, and lower wage rates. They are largest for women who have children earlier in life (Kahn et al. 2014) and for highly paid women (England et al. 2016).

The motherhood penalty is one manifestation of the contradiction between production and reproduction generated by an economic system which privileges profits above care. The capitalist system separates production from reproduction, privileging the former over the latter. In this system, employers' profit maximization objective threatens the necessary conditions of their existence (labor). An employer's survival (ability to make a profit) is heavily reliant on labor while at the same time the demands of waged work (and of capital accumulation in general) undermine labor's ability to reproduce itself (Engels 1845, Fraser 2017). This contradiction within the system places reproduction in contradiction to production, and thus manifests as tradeoffs between the two. The omnipresent tradeoff women face between family and career is a manifestation of this contradiction.

---

[7] McGinley (2016) makes a similar argument, although focuses exclusively on egg-freezing.

Some scholars have described this tradeoff women face as a double bind – a situation in one must navigate conflicting demands – the demands of waged labor with those of childrearing. This double bind is created by a conflict between the social pressure and desire to bear children and the economic realities faced by women. In the current system, the demands of waged work and those of childrearing are often at odds with each other. For example, childrearing requires flexible hours but waged work may be contingent upon long and unpredictable hours. In terms of timing their births, women must choose between two equally unsatisfactory outcomes: face the economic penalties that result from having children early in one's career or face the biological costs of delaying childbirth (Goodwin 2005).[8] In this context, women may feel pressured to delay childbirth or have fewer children to increase the likelihood of economic success – incurring reproductive costs for economic gains.

Employer subsidization of these policies likely exacerbates the pressure women face to delay childbearing or have fewer children in order to secure economic success or security.

---

[8] The costs of late childbirth often include the economic, emotional, and biological strains associated with using assisted reproductive technologies such as IVF and egg freezing (Goodwin 2005).

Indeed, assisted reproductive technologies have been marketed as a pathway to economic advancement, with the message that women can advance economically by timing their births correctly. However, many women are misinformed about the reproductive risks associated with delayed childbirth and the technologies which enable it (Tough et al. 2007, Daniluk and Koert 2012). The success rate of egg-freezing and IVF declines rapidly with age. In the case of IVF, the success rate is 51.0% for women younger than 35, but it drops to 25.1% for women aged 38-40 and falls further to 12.7% for women aged 41-42 (SART 2019). Despite the risks, media depictions of social egg-freezing disproportionately emphasize the positive aspects of it. Campo-Engelstein et al. (2018) find that Apple and Facebook's announcement to include egg-freezing in their benefits package prompted significant shifts in public discourse of social egg-freezing. After the announcement, coverage of social egg freezing shifted from focusing on drawbacks of the procedure to the benefits of it. They conclude that media depictions of egg-freezing "paint a simplistic and rosy picture that more options, especially more reproductive and economic options, automatically enhance women's autonomy" (Campo-Engelstein et al. 2018, pg. 181) This context complicates our understanding of choice in reproductive decision making, raising questions about women's agency in these decisions.

Labor markets operate in ways that not only fail to acknowledge the contributions of the reproductive economy but also disadvantage those who do reproductive work – mainly women. Employers, thus, view caring labor as a liability rather than an asset. However, labor market institutions cannot escape the fact that someone has to pay for the kids – thus, they are constructed on the basis that reproductive burdens are largely borne by women (Elson 1999). Elson (1999) warns that adaptations such as part-time work are often one-sided in that they are more designed to allow employers access to workers whose entry into the labor market is constrained by domestic responsibilities (Elson 1999). Employer subsidization of abortion and assisted reproductive technologies can be seen as one of these one-sided adaptations where instead of reorganizing the workplace around women's reproductive schedules, women are encouraged to extend their reproductive lives to better balance their reproductive and productive work. Others have come to similar conclusions though focusing only on egg-freezing.[9] Thus, we conclude that reproductive justice is not furthered by employer subsidization of abortion and assisted reproductive technologies because it reifies existing structures which incentivize women to delay childbirth and reduce their fertility.

## 7. Economic and Reproductive Inequalities

Employer subsidization of abortion and assisted reproductive technologies also detracts from reproductive justice because the policy reinforces and exacerbates existing reproductive and economic inequalities. This (perhaps) unintended consequence highlights the interconnectedness of reproductive and economic justice, something reproductive justice advocates and feminist economists have long recognized (Gammage et al. 2020). This interconnectedness is well documented in the literature on stratified reproduction, which suggests that reproductive agency is distributed along the same line as social inequality (Riley 2018). More specifically, stratified reproduction refers to the imbalances in the ability of people with different socially salient identities such as class, race, gender, and nationality to reproduce and nurture their children, and thus control their reproduction (Reiter and Ginsburg 1995, Colen 1995). The longstanding inequalities in access to reproductive health services is one example of stratified reproduction.

---

[9] Bozzaro (2018) makes an analogous argument about social egg freezing, egg-freezing for non-medical reasons. She concludes that while social egg freezing allows women to adapt to current socioeconomic constraints, it does not address the factors which lead women to delay childbirth, as these factors cannot be solved by simply extending a woman's fertility.

In the case of abortion, these inequalities stem from several decades of increasingly restrictive legislation at the state level. The shrinking geography of abortion provision carved out by legislative and religious campaigns of abortion opponents has increasingly pushed abortion out of reach for women of color and low-income women. For example, mandatory waiting periods that require women to have two face-to-face visits with an abortion provider result in an average 8.9% reduction in total abortions and a 1.5% increase in birth rates across the U.S. The effects are 2.5 times larger for Black women than White women, and three times larger for younger women compared to women in their thirties (Myers 2021a). Similarly, increases in travel distance caused by clinic closures disproportionately affect Black and young women (Myers 2021b). The introduction of COVID-related restrictions on abortion services to this legal landscape intensified the barriers that abortion seekers and providers already faced, with disproportionate impacts on Black and Hispanic abortion seekers (Wolfe and Rodgers 2022).

Access to IVF, egg freezing, and surrogacy is similarly stratified by race/ethnicity and class, largely because of the financial costs of these services. For example, out-of-pocket costs for IVF range from about $10,000 to $15,000 per treatment cycle, and only nine U.S. states have mandated insurance coverage of IVF; even those with insurance coverage still have an out-of-pocket expense of $2,000 to $3,000 (Hamilton et al. 2018). Insurance mandates to cover IVF have led to meaningful increases in the use of fertility treatments, in delayed marriage, and in the probability of having a first child for White women, but not Black women (Abramowiz 2014; Bitler and Schmidt 2012; Ohinata 2011). In fact, non-White and less-educated women are more likely to have problems with fertility but are less likely to utilize treatment for infertility (Bitler and Schmidt 2006). Egg freezing is also expensive, with an average cost per cycle of about $7,000 in the U.S., and it usually is not covered by private or public health insurance plans. These financial barriers mean that egg freezing remains out of reach except for those who are predominantly White, upper-middle-class professionals (Inhorn et al. 2018).

Not only do marginalized women face constraints that limit their access to a full range of contraceptive technologies, they may also lack the resources to be able to support their children in safe and healthy environments. As a result, marginalized women have very little control over their own reproduction. In particular, low-income women, women of color, and women with limited healthcare access have a higher frequency of reproductive events (infertility, abortion, (un)intended pregnancy, and miscarriages/stillbirths) throughout their lifetime, as well as more complexity in their reproductive lives, than their high-income, white counterparts (Johnson et al. 2023). Despite large barriers to access, the majority of women who get abortions are low-income women of marginalized social groups, a reflection of the self-reinforcing nature of reproductive and economic inequalities (Jerman et al. 2016). Angela Davis amplifies this point as follows: "when Black and Latina women resort to abortions in such large numbers, the stories they tell are not so much about their desire to be free of their pregnancy, but rather about the miserable social conditions which dissuade them from bring new lives into the world" (Davis 1983, pg.355). This argument is echoed in the results of the Turnaway Study, which finds that women who sought abortions but were denied due to restrictive laws were less likely to be employed full time and more likely to live in poverty and to require public assistance compared to women who obtained abortions (Foster et al. 2022).

Within this context, where marginalized women have little control over reproduction but where control over reproduction is highly determinant of economic success, employers are likely to gain power over these workers by offering access to technologies that allow women greater control over their fertility. Marginalized women may then be more susceptible to job lock, willing to work at companies under working conditions they otherwise would not have tolerated in order to gain more control over their reproduction. Chapin (2022) supports this assertion with evidence on the emergence of social media groups whose explicit purpose is to share information and resources with other women about how to obtain employment in order to cover fertility-enhancing services such as IVF. Facebook support groups and TikTok accounts help women navigate the pros and cons of insurance plans at different employers, mainly Starbucks and Amazon. Amazon offers employees support for at least two IVF cycles. Chapin interviewed 14 current and former Amazon-warehouse workers who had applied to Amazon exclusively for access to fertility-related benefits. Most of them reported staying in distressing working conditions for fear of losing their benefits. In another example of how reproductive inequalities and economic inequalities are self-reinforcing, one interviewee described feeling hostage, noting, "If I get fired, I cannot have a baby." This worker describes a kind of reproductive job lock where an individual's ability to reproduce is contingent on employment at a specific employer.

Employer subsidization of reproductive healthcare services can be understood as one example of how a woman's ability to control her reproduction becomes stratified by the social and economic factors that inform inequality nationally and globally. Specifically, employer subsidization disproportionately increases access for high-income women who were already best able to access to the relatively costly assisted reproductive technologies. The employers offering these benefits are disproportionately large employers employing high-income, white-collar workers (Dowling 2021, Rhia Ventures 2023). High-income women have the bargaining power and resources to incentivize employers to subsidize their reproduction. Additionally, employers usually provide these benefits through their health insurance plans, which tend to exclude part-time workers and contractors. This leaves part-time workers, contractors, the unemployed, and those out of the labor force at a disadvantage.

High-income women whose employers subsidize access to assisted reproductive technologies are likely to feel pressure to delay childbirth. These women tend to be highly educated and work in jobs requiring specialized skills and training, contributing to the steep earnings penalties they incur when they have children early in their careers. These penalties likely pressure women to delay childbirth (Doren 2019, Landivar 2020). Thus, by making delayed childbirth less costly, employer subsidization of assisted reproductive technologies increases the likelihood that women will succumb to this pressure and remain in the career fast track. Employer support may also increase the pressure to delay childbearing if the support is interpreted by employees as a message that the employer prefers that the worker defers childbearing (Mertes 2015). By offering support for these reproductive technologies, employers may signal to their employees that having children while working will be best tolerated if childbirth happens towards the end of a woman's reproductive life – after the employer has extracted as much work as possible before childrearing becomes a distraction. Since high-income women are better able to mitigate the motherhood wage penalty by delaying childbirth, these pressures are greatest for women with higher educational attainment (Doren 2019, Landivar 2020).

## 8. Conclusion and Policy Implications

Reproductive justice aims not to shift individual behaviors to adapt to unjust conditions, but rather to identify, challenge, and transform the unjust conditions in the first place. The challenge, then, is to recommend policies that will move countries towards a system which unravels the double bind, so women do not need to sacrifice economic for reproductive goals. To start, governments and employers must enact policies which redistribute the costs of childrearing so the costs are not primarily born by women (England and Folbre 1999). Proven policies to better support working parents, and especially working mothers, include the provision of paid parental leave and paid sick leave, universal free/affordable childcare, shorter work weeks, and policies that support workplace flexibility. Reducing the costs of children to women must involve moving toward a more equitable distribution of unpaid work in the home between men and women, so policies should cover both parents, not just mothers. Additionally, the public sector must implement universal healthcare policies that include access to sexual and reproductive healthcare services. Healthcare systems must also provide safe, affordable, and accessible abortion and contraceptive methods. Legally protecting abortion does not go far enough – legal protection is meaningless if it remains out of reach for many women. In countries where assisted reproductive technologies are available, the state should mandate that insurers cover infertility treatments and that these treatments are affordable.

We also advocate for policies that promote an economic environment in which children can be raised free of poverty and economic distress. Freedom to choose to have a child requires economic conditions in which having children does not lead to economic disaster. Thus, women's agency in reproductive decision-making can be bolstered with stronger poverty-reduction policies and higher minimum wages, as well as investments in schooling and healthcare. Rearing children must be regarded as productive work so poverty-reduction policies should not include employment requirements for parents. These policies can promote reproductive justice by loosening the economic constraints on women's reproductive decisions, and by increasing the ability of women to raise their children in safe and healthy environments.

While the policies suggested so far are a great start, larger systemic change is likely necessary to achieve reproductive justice. The current "double bind" that women find themselves in is a result of unjust conditions where reproduction is put in contradiction to production. Employer support for abortion and assisted reproductive technologies then acts as a band-aid, treating a symptom (misalignment of the biological clock with the corporate clock) of a deeply-rooted problem (contradiction between production and reproduction). Thus, tinkering with policies within a system that privileges profits is unlikely to completely undo this bind, although the policies we recommend will certainly loosen it. In order to unravel this bind, the economic context must be altered so that having a child does not threaten economic success. Such a change will likely require a transformation of our economic system into one that puts care rather than profits at the center.

**Figure 1 – Companies Supporting Abortion by Sector**

Panel A. Communication Services

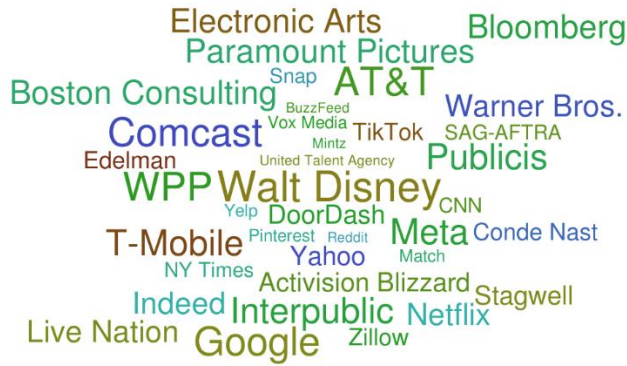

Panel B. Consumer Cyclical

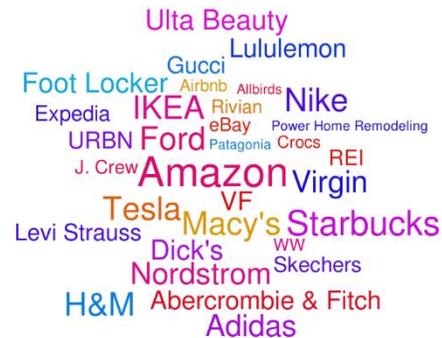

Panel C. Technology

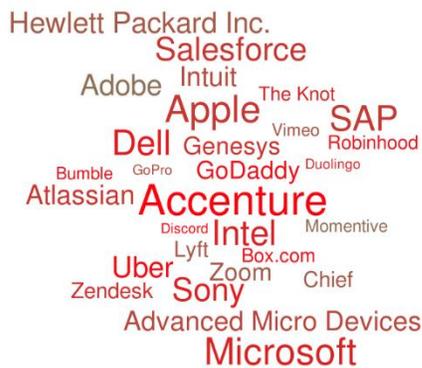

Panel D. Consumer Defensive

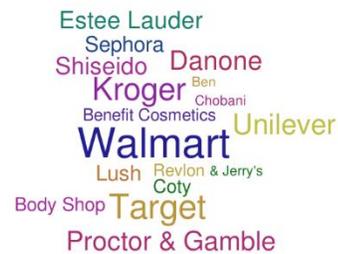

Panel E. Healthcare

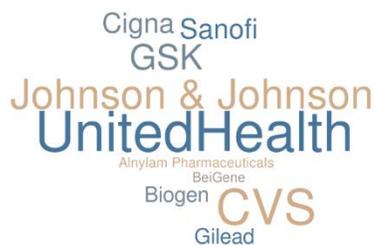

Panel F. Finance, Real Estate, Industrials

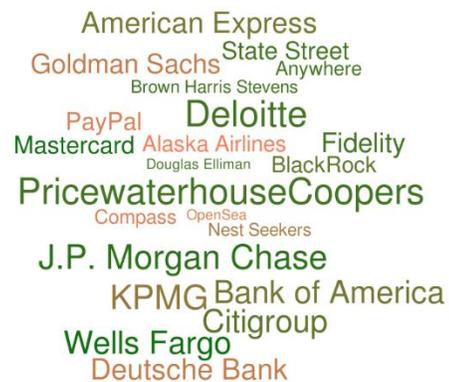

Word size correlated with # of employees. Constructed using Rhia Ventures (2023) for abortion support, and Morningstar and Google for sector and # of employees. Complete data found in Online Appendix.

**Figure 2** – Mothers' Labor Force Participation Rates by Age of Children

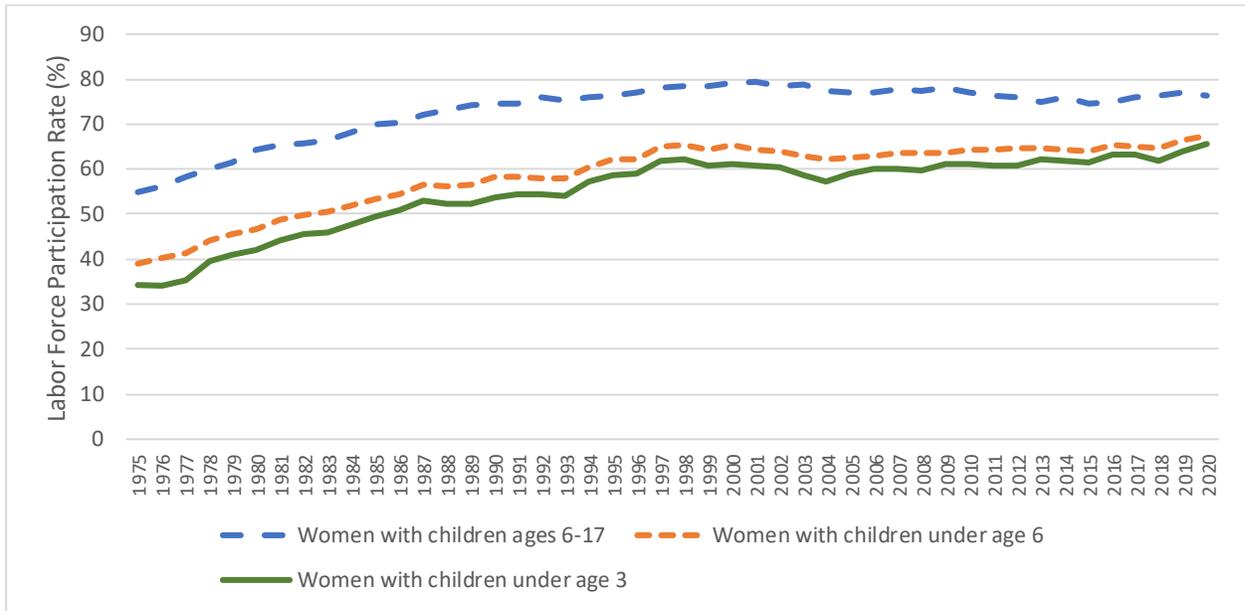

**Online Appendix:** Companies Providing Abortion Support in the U.S.

| Company name by sector | # employees | Abortion support policy or statement |
|---|---|---|
| | | |
| **Communication Services** | | |
| Activision Blizzard | 9,800 | The video game publisher added travel benefits to cover reproductive, gender-affirming, transplant health care among other services not available in the state where the employee or dependent reside. |
| AT&T, Inc. | 203,000 | The communications company stated that it will reimburse for travel costs for employees seeking abortion care more than 100 miles from where they reside. |
| Bloomberg LP | 19,000 | The parent company of Bloomberg News covers out-of-state travel for employees and dependents medical services, including abortion care, via its insurance provider where "there is no licensed provider" in the employee's state. |
| Boston Consulting Group | 21,000 | The consulting group announced that it would cover abortion-related travel expenses. |
| BuzzFeed, Inc. | 1,522 | CEO Jonah Peretti shared that the company will cover travel costs for abortion care for employees living in the 13 states with trigger bans. |
| CNN | 5,392 | CNN announced that it will cover travel costs for employees seeking abortion care. |
| Comcast Corporation | 186,000 | Comcast will cover up to $4,000 per trip, up to three times a year, with an annual cap of $10,000 for medical travel. |
| Conde Nast | 7,718 | CEO Roger Lynch shared with employees that the media company has made enhancements to its health insurance to ensure coverage of abortion care for all employees and dependents regardless of where they live, including travel and lodging costs. This benefit is also for infertility and gender affirming services. |
| The Walt Disney Company | 220,000 | The company will cover an employee's need to travel to access care, including abortion care and family planning services. |
| DoorDash | 8,600 | DoorDash confirmed that it will cover out-of-state travel costs for employees and dependents enrolled in its health plans for abortion-related care. Employees can backdate their travel costs as the company rolls out the benefit. This benefit excludes independent contractors, which includes all of its delivery drivers |
| Edelman | 6,000 | The public relations firm already covers travel for employees seeking abortion or gender-affirming care out-of-state. |
| Electronic Arts, Inc. | 12,900 | The company will expand its travel health benefits for employees and eligible dependents. |
| Google | 190,234 | Google parent company Alphabet, Inc. offers coverage for out-of-state travel for abortion care to its full-time employees enrolled in its health plans. |
| Indeed, Inc. | 12,674 | The company shared in a statement that employees enrolled in its health insurance will "continue to be reimbursed for travel expenses for covered medical procedures that are unavailable where they live." |
| Interpublic Group of Companies, Inc. | 58,500 | The company will update its health benefits to cover the cost of employees who need to travel for abortion care and other critical medical services. |
| Live Nation Entertainment, Inc. | 10,200 | The entertainment company shared on Instagram that it will cover travel expenses for employees that need to travel out-of-state for "women's health care services." |
| Match Group | 2,540 | Match CEO Shar Dubey created a fund to cover the costs of employees and their dependents who need to travel to access abortion care. |
| Meta Platforms | 86,482 | The family of social media companies shared its intention to reimburse employees who have to travel for reproductive health services. The company added the caveat that it will offer this benefit, "to the extent permitted by the law" and that they are assessing how best to provide the benefit given the legal complexities. |
| Mintz, Levin, Cohn, Ferris, Glovsky & Popeo, P.C. | 550 | Implemented compassionate leave policy for attorneys and professional staff which includes 15 consecutive days after a miscarriage, up to five days in a 12-month period after a failed surrogacy, adoption, or infertility treatment, and increased the bereavement leave up to 15 days for spouses and children and up to five days for other close relatives. |
| Netflix, Inc. | 12,800 | The streaming company will reimburse travel expenses through its health plans for employees seeking abortion care. There is a $10K lifetime max per employee or dependent per service . |
| The New York Times | 5,000 | The news outlet stated that it will expand its health benefits to "cover abortion-related travel and other procedures not available within 100 miles of an employee's home, including gender-affirming care." The company is also speaking with its unions to ensure unionized employees can also access the benefit. |
| Paramount Pictures Corporation | 22,965 | The media company's leadership sent a memo to employees committing to covering travel costs for employees enrolled in the company's health insurance. The benefit will be used available for services including abortion, miscarriage management, birth control, and other reproductive health-related services that are prohibited in the employees area. |
| Pinterest, Inc. | 3,987 | As of January 2022, Pinterest's leave policy includes four weeks of paid leave for pregnancy loss. |

| Publicis Group SA | 95,801 | During a global meeting, the company reassured staff that they will "continue to support access to reproductive healthcare for all our people throughout the U.S. -- which includes supporting our employees with travel for abortion care." |
|---|---|---|
| Reddit, Inc. | 700 | The company will provide stipends for travel expenses incurred by employees seeking procedures, including abortion care. |
| Screen Actors Guild AFTRA Health Plan | 5,250 | The SAG-AFTRA trustees approved coverage for reimbursement of travel and lodging for participants and dependents (including dependent children) who need to travel out-of-state for abortion care. |
| Snap, Inc. | 5,288 | The social media parent company informed employees the company would be reimburse up to $10,000 for travel and lodging costs related to care banned in their state, including abortion care. The coverage is for subscribers and their dependents through insurance. |
| Stagwell Group | 9,100 | The company will expand its benefits to include travel for employees to "access the nearest approved reproductive healthcare provider in a legally permissible way." |
| T-Mobile US Inc. | 75,000 | CEO Mike Sievert announced that the company expanded its existing travel and lodging coverage for care not locally available to include abortion care through its UnitedHealthCare and Premera plans. |
| TikTok | 8,424 | TikTok stated that it was in the process of updating its benefits to include ensuring employees have access to the wide range of care including family and reproductive care regardless of where they reside. |
| United Talent Agency | 1,400 | UTA introduced a new employee benefit to cover travel expenses related to seeking reproductive health services not accessible in their state of residence. |
| Vox Media | 1,900 | The media company will reimburse employees who have to travel more than 100 miles for "critical health care" up to $1,500 reimbursement for travel-related expenses. The company's union employees demanded the same benefit be available to them. |
| Warner Bros. Discovery, Inc. | 11,000 | The film company expanded its health benefits to cover travel for employees that must travel out-of-state for reproductive health care, including abortion care. |
| WPP Group, Inc. | 115,000 | CEO Mark Read sent an internal memo that WPP is "updating its benefits plan to provide funding for travel that allows consistent access to healthcare and resources, including abortion care," which will be shared in detail with employees in the next two weeks. The plan ensures confidentiality and privacy are protected at all times. |
| Yahoo, Inc. | 8,500 | The web service provider informed employees prior to the SCOTUS decision that it would cover travel costs for those who need to travel more than 100 miles to access care, |
| | | including abortion and contraceptive care. The benefit will provide reimbursement up to $5,000. |
| Yelp, Inc. | 4,400 | Announced coverage of costs for employees who must travel to access abortion care. Yelp's Chief Diversity Officer stated, "We've long been a strong advocate for equity in the workplace and believe that gender equity cannot be achieved if women's rights are restricted" . |
| Zillow, Inc. | 5,830 | The company will cover its employees travel costs up to $7,500 for those needing to travel to receive abortion or gender-affirming care. The new benefit was effective as of June 1. |
| | | |
| **Consumer Cyclical** | | |
| Abercrombie & Fitch, Co. | 31,500 | The retailer provides travel and lodging reimbursement for its associates and dependents seeking out-of-state care. |
| Adidas AG | 60,661 | The footwear company announced a new program that would cover US employees travel and lodging costs up to $10,000 through its medical plan for abortion care unavailable in state. |
| Airbnb, Inc. | 6,132 | Airbnb shared that it will continue to cover travel for employees seeking abortion care outside of their state. The company had previously committed to "support those employees whose ability to make choices about their reproductive care may be impacted by the Texas law." |
| Allbirds, Inc. | 710 | The company's leadership will support employees seeking an abortion, "Should you have to incur travel to reach a state that legally allows an abortion and you would like to exercise your right to do so, we will cover your travel cost to ensure that you can make the decision that is right for you." The coverage includes cost of a support person to travel with the employee and any childcare costs incurred during that time. |
| Amazon, Inc. | 1,541,000 | Amazon is expected to cover travel expenses for abortion-related travel for employees and dependents eligible for health care insurance, up to $4000 annually. The care must not be available via telehealth or within 100 miles of the employee's home. Amazon offers up to $10,000 annually for travel reimbursements for life-threatening conditions. The coverage does not include contract employees. |
| Crocs, Inc. | 5,770 | The footwear company announced that it will expand its benefits to include reimbursement up to $4,000 for employee travel costs who need to travel out of state for abortion care. |
| Dick's Sporting Goods, Inc. | 50,800 | The sporting goods company will provide up to $4000 in travel expense reimbursement for employees, their spouse, or dependent enrolled in their company's health plan who live in |
| | | states where abortion is restricted. The benefit will also cover one support person's travel expenses. |
| eBay, Inc. | 10,800 | The e-commerce company will reimburse employees and other covered beneficiaries who need to travel for care when not locally available and unable to be address via telehealth. The benefit is available as of June 8. |
| Expedia Group, Inc. | 14,800 | The travel company stated that it would help cover the costs of travel for employees who need "health care procedures" not provided locally. |
| Foot Locker Retail, Inc. | 49,933 | The footwear company will reimburse employees enrolled in the insurance plan for travel for medical care including family planning. |
| Ford Motors Company | 173,000 | The company shared that employees that have a health savings account (HSA) could use their HSA to get reimbursed for travel costs for medical care. |
| Gucci, Inc. | 17,157 | The luxury brand will provide travel reimbursement to employees in the US that need to access care outside of their home state. |
| H&M | 155,000 | The clothing company shared that it will cover travel-related expenses for employees that live in states with abortion restrictions. |
| IKEA US | 231,000 | Ikea announced that it is expanding its benefits to include reimbursement for eligible travel expenses for abortion, fertility, gender affirming care, and bariatric surgery "when it is unavailable within a reasonable distance of a co-worker's home or in their state of residence." |
| J. Crew Group, Inc. | 9,400 | The company's CEO stated, "While we are still navigating this new legal reality, we are prepared to use whatever lawful means possible to assist our employees who need special travel to access healthcare." |
| Levi Strauss & Co. | 18,000 | Levi Strauss issued a statement offering employees who participate in the company's health plans a travel benefit to access abortion care out of state; part-time employees can be reimbursed. |
| Lululemon Athletica, Inc. | 29,000 | The clothing brand expanded its support for reproductive rights by donating $500,000 to the Center for Reproductive Rights and continuing its support for the Black Women's Health Imperative. |
| Macy's, Inc. | 88,857 | The department store stated that it will expand its benefits coverage to provide travel reimbursement for employees to receive medical care they need. |
| Nike, Inc. | 79,100 | In its statement, Nike shared that it covers travel and lodging expenses where care is unavailable near an employee's home. |
| Nordstrom, Inc. | 60,000 | The fashion retailer expanded its health benefit to include coverage for travel to ensure its "employees have continued access to the healthcare they need." |
| Patagonia, Inc. | 1,000 | The outdoor brand stated that it covers the cost of health insurance for both part-time and full-time employees. The health plans cover abortion care and cover travel, lodging, and food for those employees who live in a state with restrictions. |
| Power Home Remodeling | 2,760 | The remodeling company released a statement announcing a policy, effective immediately, to reimburse employees or dependents that need to travel to receive medical care not available within 100 miles of their residence. The coverage includes "airfare, mileage/gasoline costs, tolls, hotel fees, meals expenses, childcare, and other applicable costs." The policy has a $5,000-lifetime reimbursement limit for individuals and a $10,000 family lifetime reimbursement limit. |
| REI Co-op, Inc. | 15,000 | The outdoor gear co-op will reimburse for travel and lodging for employees living in restricted states and need to travel more than 100 miles for reproductive health care. |
| Rivian Automotive | 10,422 | The carmaker committed to providing up to $5,000 employee and dependent travel expenses who need to access reproductive health care. |
| Skechers, Inc. | 11,700 | The footwear company will expand its benefits program to include reimbursement of $4,000 for travel expenses for medical care, including abortion. CEO Mark Greenberg stated, "Good corporate citizenship means equal opportunity for all employees, and as a company with employees in all 50 states, we believe it is up to us to do what we can to provide the same rights across our U.S. workforce." |
| Starbucks Corporation | 402,000 | The company announced in May that they are expanding its health care to include reimbursement for travel expenses to seek abortion or gender-affirming care when unavailable within 100 miles of the employee's or dependent's home. This only applies to employees and dependents enrolled in Starbucks health plans. |
| Tesla, inc. | 127,855 | Tesla provides coverage for travel and lodging for those seeking health care unavailable in their home state through an "expanded Safety Net program" offered since 2021. |
| Ulta Beauty | 40,500 | The company will cover travel costs for employees enrolled in its health plan who need to travel for reproductive health care services. |
| URBN | 23,000 | URBN told The Washington Post that it will cover travel for abortion so that its employees "can access the comprehensive benefits offered by our health plan, no matter where they live." |
| VF Corporation | 35,000 | CEO announced on LinkedIn "one of our highest priorities at VF is providing robust benefits and resources to our associates so you can manage your physical, financial, and |
| | | emotional well-being. As we shared a few weeks ago, our HR Benefits team has updated our national health plans to extend travel and lodging benefits for all medically necessary services that aren't available in the areas where our associates reside, including elective abortion." |
| Virgin Group Ltd. | 71,000 | Founder Richard Branson issued statement affirming his support of abortion access. |
| WW International | 7,700 | Weight Watchers will provide support for travel needed to access reproductive health care. |
| | | |
| **Consumer Defensive** | | |
| Ben & Jerry's Homemade Holdings, Inc. | 999 | Ben and Jerry's issued a statement condemning SB 8 as a racist law that denies people bodily autonomy. The company shared in an interview that it supports the Women's Health Protect Act, and that insurance coverage includes abortion care. |
| Benefit Cosmetics, LLC | 4,178 | The beauty brand committed to cover travel expenses for its employees who cannot access care where they live. |
| The Body Shop | 10,000 | The Body Shop states that as part of its priority to support its employees affected by the Dobb's decision, it will reimburse for expenses for US employees so they can access safe care where |
| Chobani, LLC | 2,001 | CEO announced on twitter that the company added travel for out-of-state care -- including "women's reproductive health services" -- to its health plan. Coverage includes transportation, lodging, and childcare costs for employees or dependents that need to travel out of state. |
| Coty | 11,012 | The beauty company shared that its US-based employees in states with restrictions or bans will be reimbursed up to $10,000 for transportation and accommodation. |
| Danone North America | 98,105 | The company updated its health benefits to include abortion-related travel. |
| The Estee Lauder Companies | 63,000 | The brand expanded its benefits program to cover travel and lodging for employees that travel to access reproductive health care when not locally available. This benefit is available to both full- and part-time employees and dependents enrolled in the company's health plan. The benefit will be effective as of August 1, 2022. |
| The Kroger Company | 420,000 | The company stated its benefits include quality, affordable health care and travel up to $4,000 for several categories of care and reproductive health services, including abortion and fertility. This benefit is available to employees enrolled in the company's health plan. |
| Lush Cosmetics Ltd. | 12,000 | The beauty company added benefits to "address a variety of medical needs, including abortion and gender-affirming care, that may require travel to different states." |
| Proctor & Gamble | 99,000 | P&G stated that effective January 1, 2023, the U.S. healthcare plans will expand coverage for travel support for travel expenses incurred to receive covered medical care when a provider is not available within a 50-mile radius. |
| Revlon | 5,800 | The beauty company committed to expanding its benefits to include up to $2,000 for travel and lodging costs for those who live in states that ban "certain services." |
| Sephora USA, Inc. | 28,540 | The company commits to cover transportation costs for employees that live in states that restrict access to abortion and need to travel to a state with access to care by October 1st. |
| Shiseido Americas | 41,931 | The conglomerate of brands expanded its employee benefits to cover travel and other expenses for employees seeking certain reproductive health procedures not available nearby. |
| Target Corporation | 450,000 | Starting in July, the company will cover travel for abortion care if the employee lives in a state where abortion is banned. |
| Unilever, Inc. | 148,000 | The beauty brand is "committed to providing our U.S. employees with comprehensive reproductive healthcare benefits to cover travel costs if care is no longer available in their home states." |
| Walmart, Inc. | 2,300,000 | Walmart shared in late August 2022 that it would cover abortions only in cases of health risk to the patient, rape, incest, miscarriage, or ectopic pregnancy. The company shared that it would provide "travel support" through its insurance plans for covered individuals if they need to travel to access care within 100 miles of their location. Walmart shared in an internal memo that it will add surrogacy support, financial support for adoption, and create a Center of Excellence for fertility care. |
| | | |
| **Financial Services** | | |
| American Express, Inc. | 64,000 | The company stated that its US health plans already cover abortion care and travel expenses. |
| Bank of America Corporation | 213,000 | After the recent SCOTUS decision, the bank shared that it will cover travel for its employees that need to go out of state for abortion care. |
| BlackRock, Inc. | 19,900 | The asset management firm announced that it will pay for travel expenses for employees that need to travel for abortion care. |
| Citigroup, Inc. | 223,444 | Citigroup announced in its 2022 proxy statement to shareholders that it will "provide travel benefits to facilitate access to adequate resources" starting in 2022 in response to the changing reproductive health laws. |
| Deloitte U.S. | 415,000 | The consulting company shared that it would cover travel for employees who need health care not available locally. |
| Deutsche Bank AG | 84,930 | The bank is updating its health policy for US employees to cover travel costs for any medical procedure, including abortion, that is unavailable within 100 miles of where the employee resides. |
| Fidelity Investments | 57,000 | Fidelity recently joined other asset managers and shared that it will cover travel for employees enrolled in the company's health plan seeking abortion care. |
| Goldman Sachs Group, Inc. | 49,100 | The investment firm will cover travel expenses for US-based employees that travel to receive abortion or gender affirming care. This benefit will be available as of July 1, 2022. |
| J.P. Morgan Chase & Co. | 288,474 | JPMorgan Chase will cover travel costs for US-based employees that need to travel more than 50 miles to get care, including abortion care. |
| KPMG, LLP | 265,000 | The multinational professional services network shared that it would cover abortion-related travel and lodging. |
| Mastercard, Inc. | 24,000 | Mastercard will cover out-of-state travel and lodging expenses for its employees seeking abortion care starting in June. |
| OpenSea | 800 | The largest marketplace for nonfungible tokens shared that it will cover travel expenses for US employees and dependents to receive "critical health care" and created a #roe-discussion Slack channel to encourage employees to support each other in response to the news. |
| PayPal Holdings, Inc. | 30,900 | PayPal's Chief Human Resources Officer Kausik Rajgopal shared that the company will reimburse employees who live in states with abortion restrictions and need to travel. |
| PricewaterhouseCoopers | 295,000 | The professional services brand shared that PWC employees could apply for financial assistance for medical expenses. |
| State Street Corp. | 41,354 | The financial institution shared that it will rework its health benefits to cover travel costs for employees seeking abortion care. |
| Wells Fargo & Company | 239,209 | The financial institution shared that it will rework its health benefits to cover travel costs for employees seeking abortion care. |
| | | |
| **Healthcare** | | |
| Alnylam Pharmaceuticals Inc. | 1,665 | The pharmaceutical company committed to covering travel for its employees seeking abortion care. The biotech company would cover all travel expenses outside the plan until the benefit is added via its insurer. |
| BeiGene | 9,000 | The company announced a new employee benefit for employees and their dependents enrolled in their health plans that will cover out-of-state travel for reproductive health services if access is restricted in their state where they live. |
| Biogen, Inc. | 9,610 | The company will add a benefit to cover travel costs for employees seeking abortion care. |
| Cigna Corp. | 73,700 | The company is expanding its travel coverage benefit for health care to include "abortion care, gender-affirming care, and behavior health services in states where access is restricted." |
| CVS Health Corp. | 300,000 | CVS said it will make out-of-state abortion care accessible to its employees. |
| Gilead Sciences, Inc. | 14,400 | In its statement, Gilead shared that it is ensuring its health plan will reimburse travel and lodging expenses for out-of-state travel for medical care related to reproductive health services. |
| GSK plc | 90,083 | The company is "committed to continuing to offer coverage for reproductive health including contraception and abortion." This includes coverage for travel and lodging as permitted by law in the US and Puerto Rico. |
| Johnson & Johnson, Inc. | 144,300 | J&J stated after the Dobbs decision that it strove to "put health within reach for the people we serve" and that "We also believe health care decisions are best determined by individuals in consultation with their health care provider." |
| Sanofi U.S. | 86,000 | The company reassured its employees it supports the right for all people to control their own bodies and that progress and equality are "intertwined, fundamental, and worth defending and protecting." |
| UnitedHealth Group | 350,000 | The health insurance agency will cover abortion-related travel. |
| | | |
| **Industrials** | | |
| Alaska Airlines | 22,918 | The company committed to continue reimbursing for travel expenses for medical procedures not available where employees live. |
| | | |
| **Real Estate** | | |
| Anywhere | 9,830 | Anywhere will reimburse up to $2,500 in travel expenses for employees; does not include brokers. |
| Brown Harris Stevens, Inc. | 2,926 | The company issued a statement that it will provide up to $4,000 for travel expenses for employees to travel to the nearest location where they can access abortion care. |

| Compass, Inc. | 4,775 | The real estate company will reimburse up to $2,500 for their employees' travel costs for seeking abortion care. The benefit will also cover companions; however, the benefit will not apply to agents since the coverage is through the health insurance plan. |
|---|---|---|
| Douglas Elliman, Inc. | 930 | The real estate company will reimburse staff and agents for travel expenses for those who are forced to seek care out of state. |
| Nest Seekers International | 1,200 | CEO Eddie Shapiro will cover "all expenses for those who ask for help." |
| | | |
| **Technology** | | |
| Accenture, Inc. | 738,000 | The consultant company will provide coverage employee travel costs who can't access health care procedures locally. |
| Adobe, Inc. | 29,239 | The company stated, "In the U.S, our healthcare plans offer consistent access to care and resources, independent of geography, which includes the coverage of abortion services and travel or lodging that may be required to obtain those services." |
| Advanced Micro Devices | 15,500 | The chipmaking company stated that its employees subscribed to US health-care plans will be reimbursed for travel and lodging for services not able to be performed in their state of residence. |
| Apple, Inc. | 164,000 | An internal memo reassured staff that it was monitoring the legal proceedings of the Texas ban, and it reminded employees that benefits included abortion care and out-of-state travel for medical if unavailable in their state. |
| Atlassian | 8,813 | The company shared that starting on June 24, only a few hours after the Dobbs decision, US employees living in states with abortion restrictions will be reimbursed for travel and lodging for both themselves and a companion if they need to go out of state to seek care. |
| Box.com | 2,172 | The company will cover employee travel and medical costs for employees seeking reproductive health care. |
| Bumble, Inc. | 900 | Created relief fund for organizations supporting the reproductive rights of people across the gender spectrum in Texas. |
| Chief | 3,300 | The company will reimburse up to $1000 for out-of-state travel expenses for employees and their family members who travel for care including reproductive and gender-affirming car. This will include child care costs. |
| Dell, Inc. | 133,000 | CEO Michael Dell sent a note to internal staff and stated in a CNBC interview that, "We generally believe that that our approach with our team members in Texas is to give them access to more health care, not less health care." |
| Discord, Inc. | 750 | The company will reimburse employees and dependents up to $5,000 for those who need to travel. |
| Duolingo, Inc. | 600 | The language-learning tech company shared in a statement that it will update its benefits to ensure all US-based employees can access reproductive health care including travel expenses for abortion care. |
| Genesys Telecommunications Laboratories, Inc. | 6,000 | CEO Tony Bates posted that the company will "pay travel expenses for any employee who chooses to travel to another state for reproductive medical procedures." |
| GoDaddy, Inc. | 6,611 | In 2021, the web host took down a site created by Texas Right to Life to collect anonymous tips on people seeking abortion care in Texas in an effort to enforce the 6-week ban. |
| GoPro, Inc. | 766 | The tech company enhanced its benefits for the full spectrum of family planning care including abortion, adoption, fertility services, prenatal and postpartum support, and surrogacy. |
| Hewlett Packard Enterprise Company | 60,200 | HPE shared that their companies would cover the costs of out-of-state travel for Texas-based employees who need an abortion. Director of issues management and policy communications stated, "HPE's medical plan allows participants to obtain care out of state, including abortion, and will cover lodging costs depending on distance traveled." |
| Hewlett Packard, Inc. | 58,000 | Hewlett Packard funds travel costs, and some lodging expenses. |
| Intel Corporation | 131,900 | The technology company said that it will continue to provide resources "for those who need to travel for save, timely health care." |
| Intuit, Inc. | 17,300 | The company stated it supports its employees' access to "comprehensive health care -- no matter where they live." The company committed to continuing to support its employees access to the full range of health care. |
| The Knot Worldwide | 1,700 | The Knot Worldwide, which includes The Bump, will provide reimbursement for transportation and travel costs to seek care that "is not accessible within a reasonable distance from an employee's home." |
| Lyft, Inc. | 5,064 | Lyft announced a $1 million donation to Planned Parenthood Federation of America to help reduce transportation as a barrier to accessing health care. |
| Microsoft Corporation | 221,000 | Microsoft will cover travel expenses for employees in the U.S. The company already covers abortion and gender-affirming care. |
| Momentive | 1,600 | The tech company will cover employee and dependent travel costs for abortion, infertility, and gender-affirming care. |

Electronic copy available at: https://ssrn.com/abstract=4739435

| | | |
|---|---|---|
| Robinhood | 2,400 | The stock-trading app will cover employee and dependent travel up to $5,000 in travel expenses for reproductive health care. |
| Salesforce, Inc. | 73,541 | Salesforce offered to pay for travel to access abortion care and/or relocate its employees and their immediate family if they have concerns about access to abortion care in their state . In September 2021, Salesforce messaged its employees offering to help relocate employees out of Texas in response to the 6-week abortion ban taking effect. |
| SAP | 111,961 | The software company expanded its US health care coverage to include out-of-state travel for employees who cannot access care they need. SAP emphasized the need to ensure "safe and consistent access to basic health services for all, including reproductive care." |
| Sony Group Corporation | 108,900 | US-based Sony employees receive reimbursement for travel if necessary to access care, including reproductive health care, under its health plan. |
| Uber Technologies, Inc. | 32,600 | CEO Dara Khosrowshahi tweeted that Uber would cover legal fees of drivers sued under SB8. |
| Vimeo | 1,219 | The video hosting, sharing, and services platform provider instituted new policies to ensure access to abortion care. CEO Anjali Sud shares that effective immediately costs related to travel and lodging will be reimbursed through their medical plan for anyone needing to travel out of state for care. |
| Zendesk, Inc. | 5,860 | In anticipation of the SCOTUS decision, the company is offering up to $3000/year to cover travel expenses for employees, spouses, or dependents who need to travel for reproductive health care, including abortion care. |
| Zoom | 8,422 | The videoconferencing company already include reproductive health care and travel coverage for more than 100 miles from home for medical care. |

Source: Rhia Ventures for abortion polices; Morningstar and Google search for sector and #employees. Data extracted Feb. 9, 2023.